\documentclass[12pt]{article}
\usepackage{times}
\usepackage{geometry}
\usepackage[utf8]{inputenc}
\usepackage{enumitem,amssymb}
\usepackage{ragged2e}
\newlist{thematic}{itemize}{8}
\setlist[thematic]{label=$\square$}
\usepackage{pifont}
\newcommand{\ANLHEP}{HEP Division, Argonne National Laboratory, Lemont, IL 60439, USA}
\newcommand{\APC}{Laboratoire Astroparticule et Cosmologie (APC), CNRS/IN2P3, Universit\'e Paris Diderot, 10, rue Alice Domon et Léonie Duquet, 75205 Paris Cedex 13, France}

\newcommand{\BenGurion}{Department of Physics, Ben-Gurion University, Be'er Sheva 84105, Israel}
\newcommand{\BNL}{Brookhaven National Laboratory, Upton, NY 11973}

\newcommand{\BU}{Boston University, Boston, MA 02215}

\newcommand{\Caltech}{California Institute of Technology, Pasadena, CA 91125}

\newcommand{\Carnegie}{The Observatories of the Carnegie Institution for Science, 813 Santa Barbara St., Pasadena, CA 91101, USA}
\newcommand{\Cavendish}{Astrophysics Group, Cavendish Laboratory, J.J.Thomson Avenue, Cambridge, CB3 0HE, UK}

\newcommand{\CPPM}{Aix Marseille Univ, CNRS/IN2P3, CPPM, Marseille, France}
\newcommand{\CEADAP}{D\'epartement d’Astrophysique, CEA Saclay DSM/Irfu, 91191 Gif-sur-Yvette, France}

\newcommand{\CfA}{Harvard-Smithsonian Center for Astrophysics, MA 02138}

\newcommand{\CITA}{Canadian Institute for Theoretical Astrophysics, University of Toronto, Toronto, ON M5S 3H8, Canada}

\newcommand{\CMUCosmo}{Department 
of Physics, McWilliams Center for Cosmology, Carnegie Mellon University}

\newcommand{\Cornell}{Cornell University, Ithaca, NY 14853}

\newcommand{\daa}{Department of Astronomy and Astrophysics, University of Toronto, ON, M5S3H4}
\newcommand{\damtp}{DAMTP, Centre for Mathematical Sciences, Wilberforce Road, Cambridge, UK, CB3 0WA}

\newcommand{\dunlap}{Dunlap Institute for Astronomy and Astrophysics, University of Toronto, ON, M5S3H4}
\newcommand{\Durham}{Department of Physics, Lower Mountjoy, South Rd, Durham DH1 3LE, United Kingdom}

\newcommand{\EPFL}{Institute of Physics, Laboratory of Astrophysics, Ecole Polytechnique Fédérale de Lausanne (EPFL), Observatoire de Sauverny, 1290 Versoix, Switzerland}
\newcommand{\ETH}{ETH Zurich, Institute for Particle Physics, 8093 Zurich, Switzerland}
\newcommand{\FNAL}{Fermi National Accelerator Laboratory, Batavia, IL 60510}
\newcommand{\FQAUB}{Dept. de F\' isica Qu\` antica i Astrof\' isica, Universitat de Barcelona, Mart\' i i Franqu\` es 1, E08028 Barcelona, Spain}

\newcommand{\GSFC}{Goddard Space Flight Center, Greenbelt, MD 20771 USA}

\newcommand{\HarvardPhys}{Department of Physics, Harvard University, Cambridge, MA 02138, USA}

\newcommand{\HKUST}{The Hong Kong University of Science and Technology, Hong Kong SAR, China}

\newcommand{\IAS}{Institute for Advanced Study, Princeton, NJ 08540}

\newcommand{\ICC}{ICC, University of Barcelona, IEEC-UB, Mart\' i i Franqu\` es, 1, E08028 Barcelona, Spain}
\newcommand{\ICCD}{Institute for Computational Cosmology, Department of Physics, Durham University, South Road, Durham, DH1 3LE, UK}
\newcommand{\ICE}{Institute of Space Sciences (ICE, CSIC), Campus UAB, Carrer de Can Magrans, s/n, 08193 Barcelona, Spain}

\newcommand{\ICTP}{International Centre for Theoretical Physics, Strada Costiera, 11, I-34151 Trieste, Italy}

\newcommand{\IFPU}{IFPU - Institute for Fundamental Physics of the Universe, Via Beirut 2, 34014 Trieste, Italy}
\newcommand{\IFT}{Instituto de Fisica Teorica UAM/CSIC, Universidad Autonoma de Madrid, 28049 Madrid, Spain}
\newcommand{\IFUNAM}{IFUNAM - Instituto de F\'{i}sica, Universidad Nacional Aut\'onoma de M\'etico, 04510 CDMX, M\'exico}

\newcommand{\INFN}{INFN – National Institute for Nuclear Physics, Via Valerio 2, I-34127 Trieste, Italy}

\newcommand{\INFNRM}{Istituto Nazionale di Fisica Nucleare, Sezione di Roma, 00185 Roma, Italy}

\newcommand{\ioa}{Institute of Astronomy, University of Cambridge,Cambridge CB3 0HA, UK}

\newcommand{\IPMU}{Kavli Insitute for the Physics and Mathematics of the Universe (WPI), University of Tokyo, 277-8583 Kashiwa , Japan}
\newcommand{\IPNL}{Universit\'e de Lyon, F-69622, Lyon, France; Universit\'e de Lyon 1, Villeurbanne; CNRS/IN2P3, Institut de Physique Nucl\'eaire de Lyon}
\newcommand{\IRFU}{IRFU, CEA, Universit\'e Paris-Saclay, F-91191 Gif-sur-Yvette, France}
\newcommand{\ITFA}{Institute for Theoretical Physics, University of Amsterdam, Science Park 904, 1098 XH Amsterdam, The Netherlands}

\newcommand{\JHU}{Johns Hopkins University, Baltimore, MD 21218}

\newcommand{\JPL}{Jet Propulsion Laboratory, California Institute of Technology, Pasadena, CA, USA}
\newcommand{\KASSI}{Korea Astronomy and Space Science Institute, Daejeon 34055, Korea}
\newcommand{\kavli}{Kavli Institute for Cosmology, Cambridge, UK, CB3 0HA}
\newcommand{\KIAS}{School of Physics, Korea Institute for Advanced Study, 85 Hoegiro, Dongdaemun-gu, Seoul 130-722, Korea}

\newcommand{\KIPAC}{Kavli Institute for Particle Astrophysics and Cosmology, Stanford 94305}

\newcommand{\KSU}{Kansas State University, Manhattan, KS 66506}

\newcommand{\LBL}{Lawrence Berkeley National Laboratory, Berkeley, CA 94720}

\newcommand{\LLNL}{Lawrence Livermore National Laboratory, Livermore, CA, 94550}

\newcommand{\LPNHE}{Sorbonne Universit\'e, Universit\'e Paris Diderot, CNRS/IN2P3, Laboratoire de Physique Nucl\'eaire et de Hautes Energies, LPNHE, 4 Place Jussieu, F-75252 Paris, France}

\newcommand{\MIT}{Massachusetts Institute of Technology, Cambridge, MA 02139}

\newcommand{\NAOC}{National Astronomical Observatories, Chinese Academy of Sciences, PR China}
\newcommand{\NCBJ}{National Center for Nuclear Research, Ul.Pasteura 7,Warsaw, Poland}

\newcommand{\NOAO}{National Optical Astronomy Observatory, 950 N. Cherry Ave., Tucson, AZ 85719 USA}

\newcommand{\OSU}{The Ohio State University, Columbus, OH 43212}
\newcommand{\OU}{Department of Physics and Astronomy, Ohio University, Clippinger Labs, Athens, OH 45701, USA}

\newcommand{\Oxford}{The University of Oxford, Oxford OX1 3RH, UK}

\newcommand{\PI}{Perimeter Institute, Waterloo, Ontario N2L 2Y5, Canada}
\newcommand{\Pitt}{University of Pittsburgh and PITT PACC, Pittsburgh, PA 15260}

\newcommand{\Port}{Institute of Cosmology \& Gravitation, University of Portsmouth, Dennis Sciama Building, Burnaby Road, Portsmouth PO1 3FX, UK}
\newcommand{\Princeton}{Princeton University, Princeton, NJ 08544}

\newcommand{\QMUL}{Queen Mary University of London, Mile End Road, London E1 4NS, United Kingdom}

\newcommand{\RomaS}{Dipartimento di Fisica, Universit\`{a} La Sapienza, P. le A. Moro 2, Roma, Italy}

\newcommand{\SCIPP}{University of California at Santa Cruz, Santa Cruz, CA 95064}
\newcommand{\Sejong}{Department of Physics and Astronomy, Sejong University, Seoul, 143-747, Korea}

\newcommand{\SHAO}{Shanghai Astronomical Observatory (SHAO), Nandan Road 80, Shanghai 200030, China}
\newcommand{\Siena}{Siena College, 515 Loudon Road, Loudonville, NY 12211, USA}
\newcommand{\SISSA}{SISSA - International School for Advanced Studies, Via Bonomea 265, 34136 Trieste, Italy}

\newcommand{\SMU}{Southern Methodist University, Dallas, TX 75275}

\newcommand{\Stanford}{Stanford University, Stanford, CA 94305}
\newcommand{\StonyBrook}{Stony Brook University, Stony Brook, NY 11794}
\newcommand{\STSCI}{Space Telescope Science Institute, Baltimore, MD 21218}

\newcommand{\SussexAstronomy}{Astronomy Centre, School of Mathematical and Physical Sciences, University of Sussex, Brighton BN1 9QH, United Kingdom}
\newcommand{\Syracuse}{Syracuse University, Syracuse, NY 13244}
\newcommand{\Tamu}{Texas A\&M University, College Station, TX 77843 }

\newcommand{\TIFR}{Tata Institute of Fundamental Research, Homi Bhabha Road, Mumbai 400005 India}

\newcommand{\UAS}{Department of Astronomy/Steward Observatory, University of Arizona, Tucson, AZ  85721}
\newcommand{\UAM}{Universidad Aut\'onoma de Madrid, 28049, Madrid, Spain}

\newcommand{\UCB}{Department of Astronomy, University of California Berkeley, Berkeley, CA 94720, USA}
\newcommand{\UCBP}{Department of Physics, University of California Berkeley, Berkeley, CA 94720, USA}
\newcommand{\UCBSSL}{Space Sciences Laboratory, University of California Berkeley, Berkeley, CA 94720, USA}

\newcommand{\UCL}{University College London, WC1E 6BT London, United Kingdom}

\newcommand{\UCSD}{University of California San Diego, La Jolla, CA 92093}
\newcommand{\UFL}{University of Florida, Gainesville, FL 32611}

\newcommand{\UGTO}{Divisi\'on de Ciencias e Ingenier\'ias, Universidad de Guanajuato, Le\'on 37150, M\'exico}

\newcommand{\Umich}{University of Michigan, Ann Arbor, MI 48109}
\newcommand{\UMN}{University of Minnesota, Minneapolis, MN 55455}

\newcommand{\UNIPD}{Dipartimento di Fisica e Astronomia ``G. Galilei'',Universit\`a degli Studi di Padova, via Marzolo 8, I-35131, Padova, Italy}
\newcommand{\UNM}{University of New Mexico, Albuquerque, NM 87131}

\newcommand{\UoM}{Jodrell Bank Center for Astrophysics, School of Physics and Astronomy, University of Manchester, Oxford Road, Manchester, M13 9PL, UK}
\newcommand{\UPenn}{Department of Physics and Astronomy, University of Pennsylvania, Philadelphia, Pennsylvania 19104, USA}
\newcommand{\UR}{Department of Physics and Astronomy, University of Rochester, 500 Joseph C. Wilson Boulevard, Rochester, NY 14627, USA}

\newcommand{\UTD}{University of Texas at Dallas, Texas 75080}

\newcommand{\Utah}{University of Utah, Department of Physics and Astronomy, 115 S 1400 E, Salt Lake City, UT 84112, USA}

\newcommand{\UWaterloo}{Department of Physics and Astronomy, University of Waterloo, 200 University Ave W, Waterloo, ON N2L 3G1, Canada}

\newcommand{\VSI}{Van Swinderen Institute for Particle Physics and Gravity, University of Groningen, Nijenborgh 4, 9747~AG~Groningen, The~Netherlands}

\newcommand{\WCA}{Centre for Astrophysics, University of Waterloo, Waterloo, Ontario N2L 3G1, Canada}

\newcommand{\WVU}{CSEE, West Virginia University, Morgantown, WV 26505, USA}
\newcommand{\WVUGWAC}{Center for Gravitational Waves and Cosmology, West Virginia University, Morgantown, WV 26505, USA}
\newcommand{\Wyoming}{Department of Physics and Astronomy, University of Wyoming, Laramie, WY 82071, USA}
\newcommand{\Yale}{Department of Physics, Yale University, New Haven, CT 06520}

\usepackage{url}
\usepackage{soul}
\usepackage[dvipsnames]{xcolor}
\usepackage{amssymb, amsmath, epsfig}
\usepackage{hyperref}
\usepackage{tabularx}
\usepackage{subfigure}
\usepackage{epstopdf}
\usepackage{comment}
\usepackage{floatrow}
\newfloatcommand{capbtabbox}{table}[][\FBwidth]

\usepackage{blindtext}

\DeclareGraphicsExtensions{.ps}
\begin{document}

\raggedright
\huge
Astro2020 Science White Paper \linebreak

Inflation and Dark Energy from spectroscopy at $z > 2$  \linebreak
\normalsize

\noindent \textbf{Thematic Areas:} Cosmology and Fundamental Physics
 \linebreak

\textbf{Principal Authors:} \\ 
Name: Simone Ferraro\\
Institution:  Lawrence Berkeley National Laboratory, One Cyclotron Road, Berkeley, CA 94720, USA\\
Email: sferraro@lbl.gov
 \linebreak
 
Name: Michael J. Wilson\\
Institution:  Lawrence Berkeley National Laboratory, One Cyclotron Road, Berkeley, CA 94720, USA\\
Email: mjwilson@lbl.gov
 \linebreak
 
\textbf{Co-authors / Endorsers:}\\
Muntazir Abidi$^{1}$, 
David Alonso$^{2}$, 
Behzad Ansarinejad$^{3}$, 
Robert Armstrong$^{4}$, 
Jacobo Asorey$^{5}$, 
Arturo Avelino$^{6}$, 
Carlo Baccigalupi$^{7,8,9}$, 
Kevin Bandura$^{10,11}$, 
Nicholas Battaglia$^{12}$, 
Chetan Bavdhankar$^{13}$
% ...

%%% 

 $^{1}$ \damtp \\
$^{2}$ \Oxford \\
$^{3}$ \Durham \\
$^{4}$ \LLNL \\
$^{5}$ \KASSI \\
$^{6}$ \CfA \\
$^{7}$ \SISSA \\
$^{8}$ \IFPU \\
$^{9}$ \INFN \\
$^{10}$ \WVU \\
$^{11}$ \WVUGWAC \\
$^{12}$ \Cornell \\
$^{13}$ \NCBJ \\
% ...

\newpage

%%%%%%%%%%%%%%%%%%%%%%%%%%%%%%%%%%%%%%%%%%%%%%%%%%%%%%%%%%%%%%%%

\justify
The expansion of the Universe is understood to have accelerated during two epochs: in its very first moments during a period of `Inflation' and much more recently, at $z < 1$, when Dark Energy is hypothesized to drive cosmic acceleration. The undiscovered mechanisms behind these two epochs represent some of the most important open problems in fundamental physics.

Most of the processes involved during Inflation impact observations on the very largest spatial scales \cite{Bartolo:2004if, Alvarez:2014vva}. Traditionally, these have been accessed through observations of the Cosmic Microwave Background (CMB). While very powerful, the CMB originates from a 2D surface and the finite number of modes that it contains will largely be measured by experiments over the next decade.\footnote{Cosmologically relevant modes of CMB temperature anisotropies have been measured to the cosmic-variance limit by Planck \cite{Akrami:2018vks} and upcoming or proposed experiments will achieve the same for polarization \cite{Ade:2018sbj,Abazajian:2016yjj,Hanany:2019lle}.} Observations of large 3D volumes with large-scale structure (LSS) access similar scales and will dramatically increase the number of available modes. For example, LSS observations in the range $2 \lesssim z \lesssim 5$  can more than triple the volume surveyed at $z \lesssim 2$, and, together with the sufficiently high galaxy number in this interval, strongly motivates a future spectroscopic survey that exploits this opportunity. 
In addition, tomography allows mapping the growth of structure with redshift, which provides robust constraints on Dark Energy and neutrino masses while relaxing restrictive assumptions such as a power-law primordial power spectrum \cite{dePutter:2014hza}.

Finally, cross-correlation with external tracers, such as CMB lensing, Intensity Mapping or the Lyman-$\alpha$ forest, immunises the constraints to the systematics that make measurement challenging and further improves the precision through `sample variance cancellation' \cite{Schmittfull:2017ffw, Munchmeyer:2018eey, Chen:2018qiu} and degeneracy breaking.

\section{Science Case}
\label{Sec:Science}
\textbf{\textit{Inflation}} 
Simple theories of inflation, involving a single non-interacting field, predict that the primordial fluctuations are extremely close to Gaussian distributed \cite{Maldacena:2002vr, Creminelli:2004yq}.  However, very large classes of inflationary models produce levels of non-Gaussianity that are detectable by the next generation of spectroscopic surveys \cite{Bartolo:2004if}. Measurements of primordial non-Gaussianity probe the dynamics and field content of the very early Universe, at energy scales far above particle colliders. Deviations from Gaussianity leave a particular imprint on the galaxy three-point correlation function or bispectrum \cite{Karagiannis:2018jdt} (and of the CMB), and can also produce a characteristic scale-dependence in the galaxy bias \cite{Dalal:2007cu}. Depending on the physical process responsible for these deviations from Gaussianity, different configurations in the three-point function are generated. These are typically described by a number of dimensionless parameters, 
$f_{NL}$ \cite{Komatsu:2001rj}, and common examples include the local, equilateral and orthogonal types. The local type is generically produced in multi-field inflation, while the equilateral type often indicates self-interaction of the inflaton.

Pushing the observational frontier to the threshold typically expected from `non-minimal' inflation ($f_{NL}\gtrsim 1$, see \cite{Alvarez:2014vva}) provides a compelling opportunity for future large-scale structure surveys.  In summary, capturing the full picture of inflation requires measuring primordial non-Gaussianity to an unprecedented level, complementing the search for primordial gravitational waves and informing us about the Universe's first moments.

\newpage
\textbf{\textit{Dark Energy}}
Many theories have been put forward to explain the late time cosmic acceleration. They range from a cosmological constant to some dynamical forms of Dark Energy or modification to General Relativity on large scales \cite{Clifton:2011jh, Mortonson:2013zfa}. 
By mapping expansion and growth at $z > 1.5$ -- deep into matter domination -- we can ease parameter 
degeneracies, better constrain potential theories of Dark Energy, and test posited modifications to General Relativity, e.g.\ by comparing measurements of growth to the amplitude of gravitational lensing of the CMB. 

%%%%%%%%%%%%%%%%%%%%%%%%%%%%%%%
\textbf{\textit{Curvature}} 
A measurement of the global value of the Universe's curvature can potentially have important implications for Inflation.  Slow-roll eternal inflation predicts $|\Omega_K| < 10^{-4}$, while false-vacuum models would be ruled out by a measurement of $\Omega_K < -10^{-4}$ \cite{Leonard:2016evk,Kleban:2012ph}. Moreover, the current bound $\Omega_K < 2 \times 10^{-3}$ \cite{Akrami:2018vks} relies on the strong assumption that Dark Energy is a cosmological constant. If this is relaxed, large degeneracies with the time evolution of Dark Energy arise, significantly degrading the constraints on both.  Measurements at high redshift can break this degeneracy and, at the same time, approach the threshold $\sigma(\Omega_{K}) \approx 10^{-4}$ that is crucial for a better understanding of Inflation \cite{Denissenya:2018zcv}.

%%%%%%%%%%%%%%%%%%%%%%%%%%%%%%%%%%
\textbf{\textit{Neutrino Masses}} Massive neutrinos suppress the growth of structure on small scales in a time-dependent manner \cite{Lesgourgues:2006nd}.
Measuring the amplitude of structure over a long lever-arm in redshift, $z \sim 0 - 5$, better constrains the neutrino masses and breaks important degeneracies with the time evolution of Dark Energy and the primordial power spectrum \cite{Allison:2015qca, Yu:2018tem}. 

%%%%%%%%%%%%%%%%%%%%%%%%%%%%%%%%%%%%
\subsection{High-$z$ Lyman-break galaxies and Lyman-$\alpha$ emitters}
Lyman-break galaxies are young, star forming galaxies that comprise the majority population at $z>1.5$.  Their characteristic spectral energy density exhibits a sharp drop in the optical flux blue-wards of the redshifted Lyman limit, $(1+z) \times 912$\AA, due to absorption by neutral hydrogen, in an otherwise shallow $F_{\nu}$ spectrum.  As such, they are efficiently selected with a search for galaxies bright in a detection band, $m_{UV}$ -- chosen to correspond to the rest-frame UV for ease -- but otherwise undetected in all bluer filters (see Refs. \cite{Giavalisco02, Shapley11b} for reviews).  In this manner, convenient target populations (BX, $u$-dropouts, $g$-dropouts and $r$-dropouts) spanning $\Delta z \simeq 1.0$ at $z \simeq 2, 3, 4$ and 5 are obtained by enforcing these criteria for increasingly red detection bands.  Selection on photometric redshift largely yields the same ends \cite{McClure18, Hathi16}.

While of great interest for providing very large populations at high redshift, to achieve the necessary spectroscopic success rate in a baseline exposure typically requires refinement to those with significant Lyman-$\alpha$ emission (LAEs).  This is traditionally achieved with narrow-band selection, but large volumes and sufficient depth are not obtainable in this manner.  Accepting some degree of increased contamination or lower completeness, broad-band selection based on the bluer continua of strong emitters has been shown to provide very encouraging results \cite{Cooke09, Stark10, Du18}.  Alternatively, one may limit oneself to only the  brightest galaxies, for which secure absorption line redshifts are also possible.

%%%%%%%%%%%%%%%%%%%%%%%%%%%%%%%%%%%%%
\subsection{Survey strategy}
We identify two galaxy surveys that we use as a baseline for forecasts of an airmass-limited 14,000 square degree survey.  Following Ref. \cite{Chen:2018qiu}, we first consider the idealised $m_{UV} = 24.5$ sample in Table \ref{Table:Opt}.  This informs what conclusions may ultimately be drawn for this science case with minimal assumptions on the required facilities and survey details.

Conversely, assuming a next generation survey speed, we posit a fiducial survey to approximate the properties shown in Table \ref{Table:PhysNum} -- assuming  completion of LSST Year 10 by first light.

%%%%%%%%%%%%%%%% 
% Idealisitic
\begin{table}[h]
\begin{center}
\small
\begin{tabular}{|c|c|c||c||c|c|c|}
\hline
$z$ & $n(z)$ [$10^{-4}\,h^3\,\mathrm{Mpc}^{-3}$] & $b(z)$ & \ \ & $z$ & $n(z)$ [$10^{-4}\,h^3\,\mathrm{Mpc}^{-3}$] & $b(z)$  \\ \hline \hline
  2.0 & 25 & 2.5 & \ \ \ \ \  & 4.0 & 1.5 & 5.8  \\  \hline
 2.5 & 12 & 3.3 &\ \ & 4.5 & 0.8 & 6.6 \\  \hline
 3.0 & 6.0 & 4.1 & \ \ & 5.0 & 0.4 & 7.4 \\  \hline
 3.5 & 3.0 & 4.9 & \ \ & \ & \ & \   \\  \hline
\end{tabular}
\vspace{-0.5cm}
\caption{\small Our `idealised' sample: a $m_{UV} = 24.5$ magnitude-limited dropout sample as defined by Ref. \cite{Chen:2018qiu}.  Here $n(z)$ and $b(z)$ correspond to the expected number density and linear galaxy bias with redshift.}
\label{Table:Opt}
\end{center}
\end{table}
%%%  Near term  %%%
\begin{table}[h]
\begin{center}
\small
\begin{tabular}{|c|c|c||c||c|c|c|}
\hline
$z$ & $n(z)$ [$10^{-4}\,h^3\,\mathrm{Mpc}^{-3}$] & $b(z)$ & \ \ & $z$ & $n(z)$ [$10^{-4}\,h^3\,\mathrm{Mpc}^{-3}$] & $b(z)$  \\ \hline \hline
 2.0 & 9.8 & 2.5 & \ \ \ \ \  & 4.0 & 1.0 & 3.5 \\  
 \hline
 3.0 & 1.2 & 4.0 & \ \ \ \ \ & 5.0 & 0.4 & 5.5 \\ 
 \hline
\end{tabular}
\vspace{-0.5cm}
\caption{\small Our `fiducial' sample achievable with next generation facilities.  The number density and galaxy bias estimates derive from Refs. \cite{Chen:2018qiu, Du18, Reddy08, Hildebrandt09, Malkan17} and \cite{Harikane18}. We find the limiting factors are efficient pre-selection of LAEs based on broad-band imaging, LSST $u$-band depth and our posited survey speed for $z=2, 3$ and 4 respectively.}
\label{Table:PhysNum}
\end{center}
\end{table}

%%%%%%%%%%%%%%%%%%%
\section{Forecasts}
\subsection{Primordial non-Gaussianity}
We follow Ref. \cite{Karagiannis:2018jdt} in order to forecast the constraints on primordial non-Gaussianity achievable with these samples. The results are shown in Table \ref{Table:fnl} when including both the power spectrum and bispectrum.
We find that local $f_{NL}$ sees the largest improvement, 
achieving $\sigma(f^{\rm local}_{NL}) \approx 0.1$ for the fiducial sample. This represents a factor of $\simeq 50$ improvement over current surveys and achieves the precision necessary for a paradigm shift in our understanding of the early Universe.  No planned survey can deliver this at such a redshift, which would be entirely complementary to lower $z$ studies \cite{Dore:2014cca}.  When including the external CMB and LSS data expected to be available by first light, the constraints on equilateral and orthogonal $f^{\rm local}_{NL}$ see additional improvements of $\sim 2$ and $3$ over current estimates. Given this achievable precision, the measurement will likely be systematics-dominated and the survey should be designed accordingly.

The importance of spectroscopy is clear from 
the sharp degradation in %$f_{NL}$ 
constraints -- a factor of 3 for both local and orthogonal, and a factor of 4 for equilateral -- if only photometric redshifts are available.
\begin{table}[h]
\begin{center}
\small
\begin{tabular}{|c|c|c|c|c|c|c|}
\hline
\shortstack{$\sigma(f_{NL})$ \\Fiducial / Idealised} & $P$ & $+ B$ & + External & \shortstack{Current \\ (Planck)}
& \shortstack{Photo-$z$  \\ degradation}\\ \hline \hline
Local & 0.75 / 0.63 & 0.11 / 0.073  & 0.11 / 0.073 & 5 & $\times 3$\\ \hline
Equilateral & -- & 43 / 23 & 23 / 18  & 43 & $\times 4$ \\ \hline 
Orthogonal & 50 / 33  & 8.8 / 5.0 & 7.5 / 4.7 & 21 & $\times 3$ \\ \hline
\end{tabular}
\vspace{-0.5cm}
\caption{\small Constraints on $f_{NL}$ for the two samples considered. $P$ denotes those derived from the power spectrum, while $+B$ includes additional constraints from the bispectrum.  External datasets include constraints on $f_{NL}$ coming from Planck \cite{Ade:2015ava}, DESI \cite{Font-Ribera:2013rwa} and Simons Observatory~\cite{Ade:2018sbj}, which are expected to complete by our first light.  In the last column, we illustrate a photo-$z$ degradation corresponding to $\sigma(z)/(1+z) = 2 \times 10^{-2}$.  %Adding a future proposed CMB-S4 type-experiment \cite{Abazajian:2016yjj} only leads to marginal improvements (at most $\sim 10 \%$ for equilateral and orthogonal).
}
\label{Table:fnl}
\end{center}
\end{table}

%%%%%%%%%%%%%%%%%%%%%%%%%%
\subsection{Dark Energy}
\begin{figure}[h]
\centering
\begin{tabular}{cc}
  \includegraphics[trim={1.1cm 0 0 0}, clip=True, width=8.5cm]{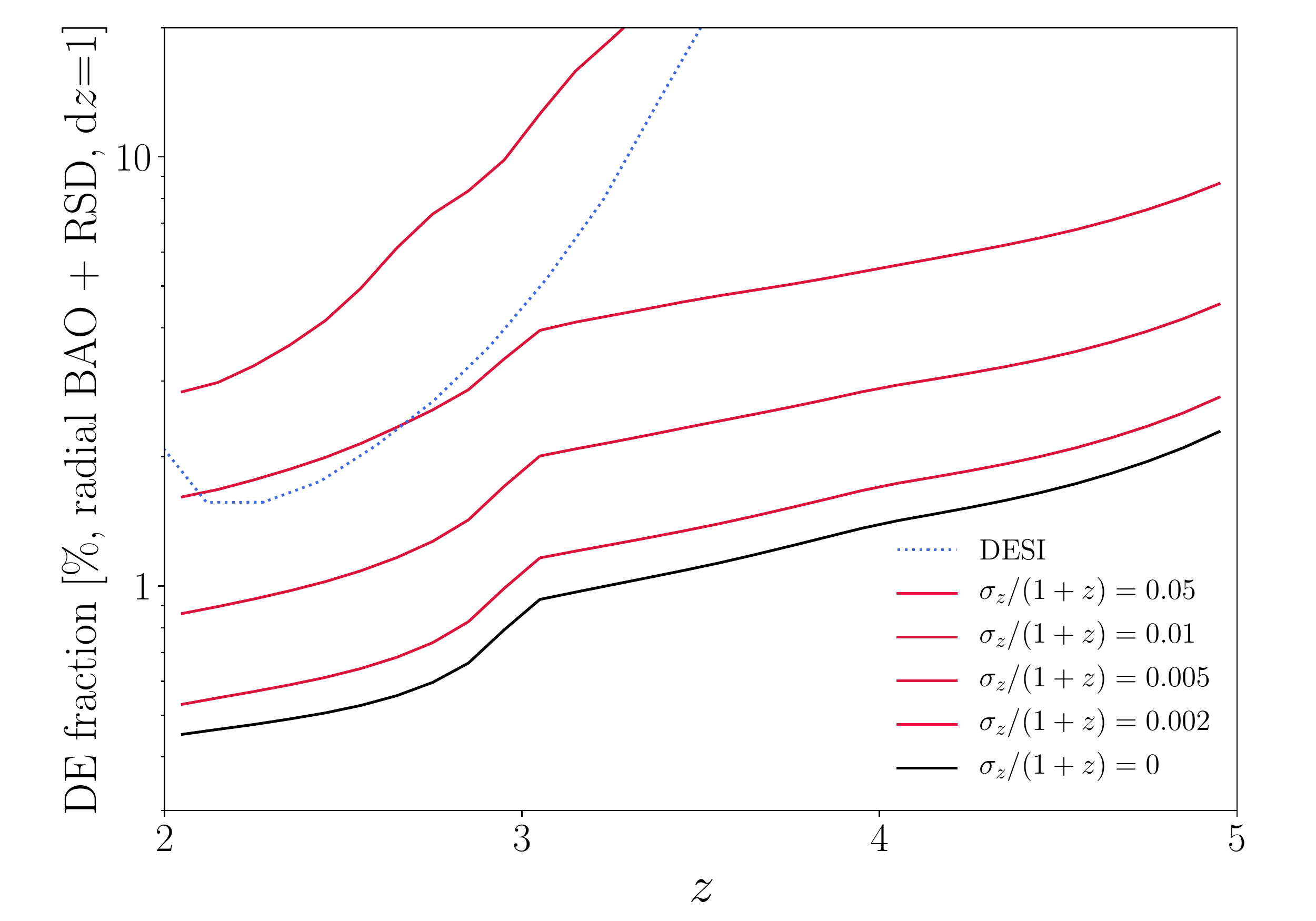} &   \includegraphics[trim={2.1cm 0 0 0}, clip=True, width=8.2cm]{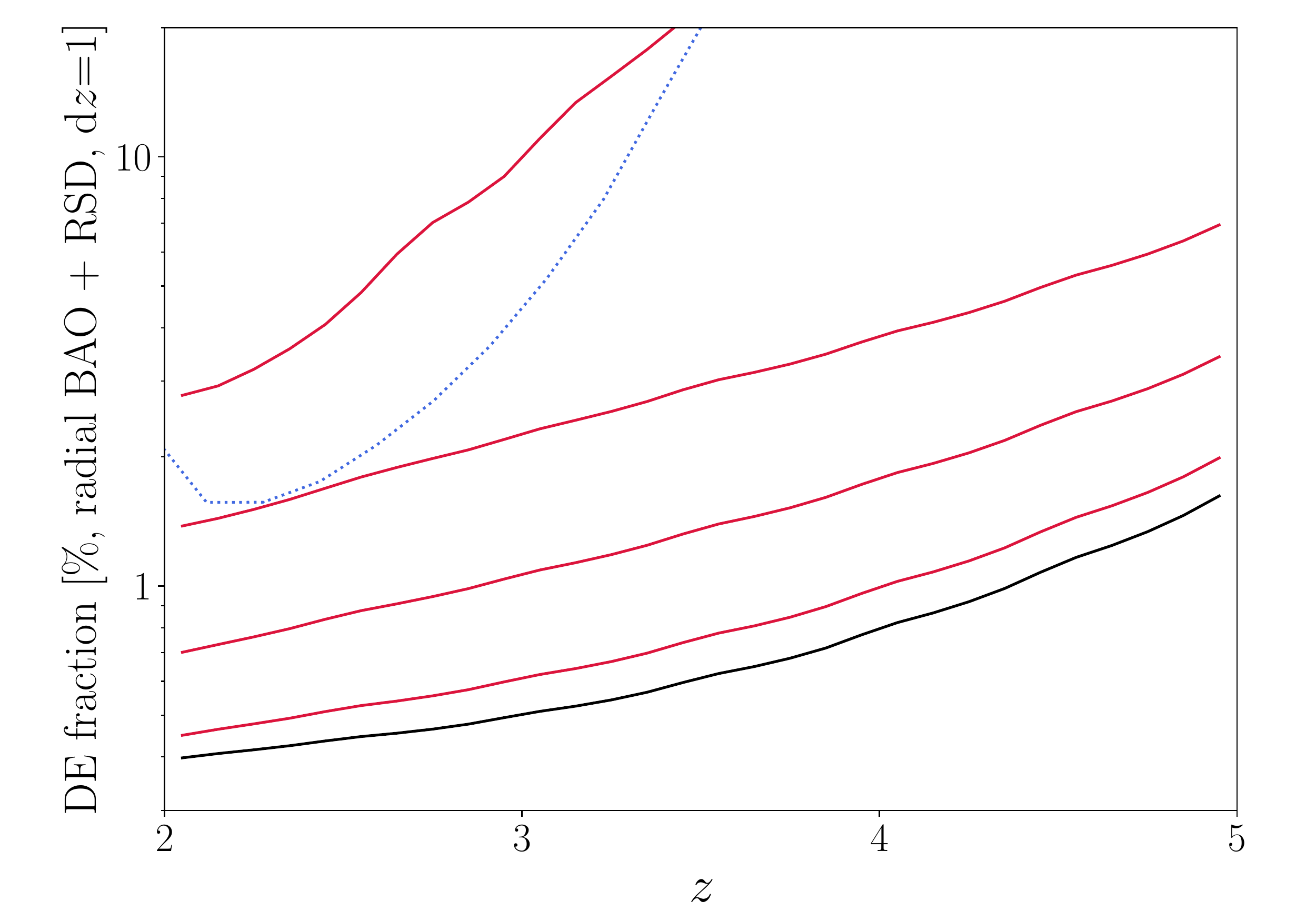} \\
\end{tabular}
\vspace{-0.5cm}
\caption{\small The absolute error on the fraction of Dark Energy $\Omega_{DE}$ at a given redshift for the fiducial (left) and idealised (right) samples.  This is obtained from a combination of radial Baryon Acoustic Oscillation (BAO) and Redshift-Space Distortions (RSD).  If Dark Energy is a cosmological constant, its fraction is forecasted to be 7\%, 3\%, 2\% and 1\% at $z = 2, 3, 4, 5$ to a very high degree of accuracy, which motivates facilities capable of challenging this prediction.}
\label{fig:DEresults}
\end{figure}

The galaxy power spectrum yields measurements of the expansion and growth rates. In turn, these can be used to infer the energy content at a particular redshift. In Figure \ref{fig:DEresults}, we show that both potential surveys constrain the fraction of Dark Energy to percent, or even sub-percent, precision to $z \sim 5$.  This would represent a tremendous increase in precision over DESI, especially for $z > 3$.
In the standard parametrization, these correspond to a Dark Energy Figure of Merit (FoM) of 398 and 441 for the fiducial and idealised samples respectively.  This is an improvement of a factor of 2.7 over DESI \cite{Font-Ribera:2013rwa} when combined with the current Planck constraints. Spectroscopy is essential in this respect, with a degradation of over $\sim 60\%$ for photometric redshifts ($\sigma(z) / (1+z) = 0.01$).
%%%%%%%%%%%%%%%%%%%%%%%%%%%

\begin{figure}
\begin{floatrow}
\capbtabbox{%
\small
  \begin{tabular}{|c|c|c|} \hline 
  Parameter & \shortstack{$\sigma$(parameter) \\ Fid./Ideal.} & DESI \\ \hline \hline
  Curvature $\Omega_K/10^{-4}$ & 6.6 / 5.2 & 12.0 \\ \hline
  Neutrinos $\sum m_{\nu}$ & 0.028 / 0.026 & 0.032 \\ \hline
  Spectral index $n_s$ & 0.0026 / 0.0026 & 0.0029 \\ \hline
  Running $\alpha_s$ & 0.003 / 0.003 & 0.004 \\ \hline
  Rel. species $N_{eff}$ & 0.069 / 0.069 & 0.078 \\ \hline
  Gravitational slip & 0.008 / 0.008  & 0.01 \\ \hline \hline
    D.E. FoM & 398 / 441 & 162\\ \hline

  \end{tabular}
}{%
  \caption{\small Forecasts on cosmological parameters from our samples, combined with Planck priors. Gravitational slip is defined as the ratio between the two potentials %$\Phi$ and $\Psi$ 
  describing the metric, in combination with a CMB experiment with map noise of 1 $\mu$K-arcmin.}%
  \vspace{0.55cm}
  \label{tab:params}
}
\ffigbox{%
  \includegraphics[trim=0.6cm 0.0cm 0.2cm 0.0cm,clip=true, width=7.5cm]{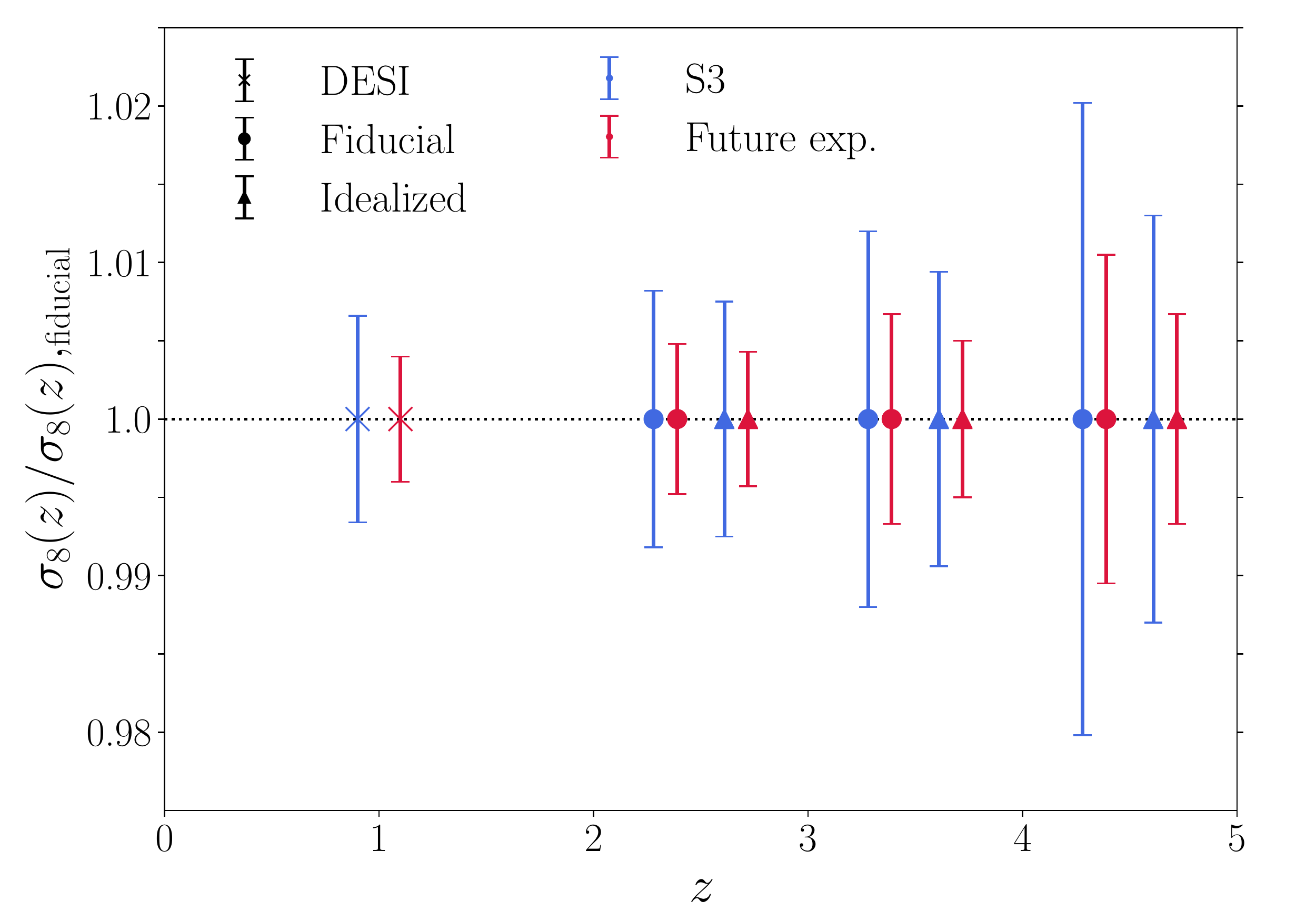}%
}{%
  \caption{\small Constraints on $\sigma_8(z)$ from cross-correlation with CMB lensing. `S3' and `Future exp.' refer to CMB experiments with map noise of 7 and 1 $\mu$K-arcmin respectively.}
  \label{fig:s8}
}
\end{floatrow}
\end{figure}

Table \ref{tab:params} shows forecasts for the (beyond) Standard Model parameters. In addition to the Dark Energy FoM, large improvements are found for the curvature $\Omega_K$ (with errors decreasing by over a factor of 2), together with the sum of neutrino masses.

While not explored in great detail here, it has been shown that cross-correlation with the CMB and Intensity mapping experiments can greatly reduce systematics and break several astrophysical and cosmological degeneracies.
As an example, Figure \ref{fig:s8} shows constraints on the amplitude of fluctuations $\sigma_8(z)$ as a function of redshift by cross-correlating CMB lensing with galaxy surveys. With this potential for synergy with future CMB surveys, we can extract sub-percent constraints on the growth that are relatively insensitive to the $z < 2$ universe and hence a powerful probe of non-standard physics.

%%%%%%%%%%%%%%%%%%%%
\section{Challenges}
Further development of efficient pre-selection of LAEs from broad-band photometry is a requirement for this case as presented.  The success of this pre-selection will largely determine the necessary facilities and achievable samples.  Some of the measurements outlined above -- especially local $f_{NL}$ -- also require complete understanding of e.g. the parent photometry and the galaxy selection function generally \cite{Alvarez:2014vva, Pullen:2012rd, Huterer:2012zs}.   Percent-level sky subtraction with fibers and exposures approaching an hour, together with mitigation of line confusion, are also technical challenges to be overcome.  Potential strategies have already been proposed and are under active study, but future surveys will require careful consideration of these points during any design phase.

%%%%%%%%%%%%%%%
\section{Conclusions}
The colossal, relatively uncharted, volume at $z>2$ and known means of efficiently selecting high-$z$ galaxies 
grants a tremendous opportunity to study the beginning and fate of our Universe, namely Inflation and Dark Energy.  We have shown potential surveys can test the %Gaussianity of the initial conditions
early Universe (Gaussianity) up to a factor of $\sim 50$ better than our current bounds %for the local shape, 
and cross the highly significant threshold of $f_{NL} \simeq 1$ that would separate single-field from multi-field models of Inflation. Such measurements would be entirely complementary to low-$z$ studies.  
This is enabled by spectroscopic redshift precision, with the lesser precision of photometric redshifts degrading these constraints by a factor of three or greater.

 Such a dataset would leave an important legacy for the science cases we have presented, together with a wealth of opportunities for the fields of galaxy formation as well as many others.
\newpage
\bibliographystyle{unsrt}
\bibliography{main}
\end{document}